\documentclass[aps, one column,prd,nofootinbib]{revtex4}
\usepackage{amsmath}
\usepackage{graphicx}
\usepackage{dcolumn}
\usepackage{bm}
\usepackage{amssymb}
\usepackage{latexsym}

\begin{document}

\title{Gravitational Casimir Effect in Inspiralling Neutron Star Binary}

\author{Jing Wang\footnote{Email address: joanwangj@mailbox.gxnu.edu.cn}}
\affiliation{School of Physical Science and Technology, Guangxi Normal University,
   Guilin, 541004, P. R. China}

\begin{abstract}
Currently, the discussions and investigations for the vacuum energy is drawing great both theoretical and experimental attention. The vacuum states of variety of fields, subject to special boundary conditions, may contribute to non-trivial macroscopic vacuum energy, i.e. the Casimir effect, which become an interdisciplinary subject and plays an important role in a variety of fields of physics. We adopt Schwinger's source theory and study the Casimir effect due to the quantization of gravitation, i.e. the gravitational Casimir effect (GCE), in inspiraling neutron star (NS) binaries with wide separation of $10^9 \rm m$. By considering gravitoelectromagnetism (GEM) arising from the spiral-in orbital motion and evaluating the contributions of GEM to the vacuum energy of gravitons radiated during the orbital decay, we demonstrate that, when the orbital separation of the binary decay a distance of $L$ in radial direction, the GEM results in a small Casimir correction to the gravitational vacuum energy, which contributes to an attractive gravitational Casimir force to the binary, in addition the gravitational force. The gravitational waves (GWs), emitted from wide inspiraling NS binaries, locate in the low-frequency band of $10^{-4}-1$ Hz. For a characteristic GW frequency of $10^{-3}$ Hz, the gravitational Casimir correction to the signals is estimated as of the order of $\sim10^{-24}$, which corresponds to a force of $10^{-20}$ N. By considering that the sensitivity of space-based gravitational wave observatory, LISA/eLISA and Taiji, can be reduced to $10^{-24}$, we would expect that LISA/eLISA and Taiji with sensitivity improvements give the powerful tool to detect GCE in the near future.
\end{abstract}

\maketitle

\section{Introduction}\label{sec:int}

The Casimir effect \cite{Plunien:1986ca, Mostepaneko:1997book} manifests the nontrivial properties of quantum fluctuations of the vacuum state, which in its simplest form is the interaction of a pair of neutral, parallel conducting plates due to the disturbance of the vacuum of the electromagnetic field. It was originally calculated \cite{Casimir1948} the shift in the vacuum energy density of electromagnetic field caused by two parallel conducting plates and was readily applied \cite{Casimir:1947kzi} to show that the classical interactions of two neutral polarizable atoms at large distances are modified by retardation effects. This was later extended by Lifshitz \cite{Lifshitz:1956zz} to forces between dielectric macroscopic bodies usually characterized by a dielectric constant. The microscopic approach to the theory of both van der Waals and Casimir forces were formulated in a unified way, non-relativistically in second order perturbation theory from the dipole-dipole interaction energy \cite{Lifshitz:1982QEDbook}. It was found that the correlation of the quantized electromagnetic field in the vacuum state is not equal to zero if we increase the distance between the two macroscopic bodies to be so large that the virtual photon emitted by an atom of one body cannot reach the second body during its lifetime. The nonzero correlated oscillations of the induced atomic dipole moments result in a Casimir force \cite{Bordag:2001qi}. The long-range interactions between polarizable systems also have been widely investigated both theoretically and experimentally \cite{Lifshitz:1956zz, Dzyaloshinskii1961, Schwinger:1977pa, 1958Phy....24..751S, 1968Natur.219.1120T}, which demonstrate that even though the quantum in nature, an important feature of the Casimir effect is that it predicts the Casimir force and gives rise to nontrivial influence between macroscopic bodies. In addition, Schwinger used the proper-time formalism for the effective action and recalculated the Casimir effect in source theoretical methods \cite{Schwinger:1951nm, Schwinger1975, Schwinger1992}, if there is no reference of quantum oscillators and zero point energy in electromagnetism.

In large-scale systems subject to the long-range gravitational interactions, the Casimir effect arises in spacetime with nontrivial topology \cite{Abe:1984ng, Appelquist:1982zs}, represent by boundaries that can be viewed as external fields. The appearance of an external field always accompanies with the effects of vacuum polarization, characterized by some nonzero vacuum energy and special boundary conditions, which has been investigated in various cases of boundary geometries and different types of field \cite{Elizalde1994book, Elizalde1995book}. The Casimir effect associated with gravitational fields at various length scales, from particle confinement to large-scale structure of the universe, has also been widely investigated theoretically and tested accurately by several experiments (\cite{Bordag:2001qi, Milton:1999ge, Bressi:2002fr} for example) in flat background. In curved spacetime \cite{Birrell1982book}, some interesting results can be obtained in special circumstances, in which the quantization procedure are performed in quite a straightforward way \cite{Calloni:2001hh, Bimonte:2008zva}. By confining inside a Casimir cavity, the possible influence of the gravitational field on the vacuum energy of a quantum field \cite{Setare:2000py, Calloni:2001mb, Caldwell:2002im, Brevik:2000zb} faces the open issue concerning the limits of validity of general relativity at small distances \cite{Mostepanenko:2000kn}. The gravitational interaction causes a small reduction in the Casimir energy of a massless scalar field confined in a rigid Casimir cavity in a slightly curved, static spacetime background \cite{Sorge:2005ed}. Based on Schwinger's effective action method \cite{Sorge:2019ldb}, it was found that the Casimir effect in a small cavity at rest in the weak gravitational field of a massive, non-rotating source suffers from a same small correction. While there is no first-order gravitomagnetic effect in the vacuum energy shift confined in the Casimir cavity, lying between two parallel massive walls that are moving in the opposite directions to each other \cite{Sorge:2009zz}.

Anyway, the macroscopic objects with given boundary conditions, e.g. Dirichlet and/or Neumann boundary conditions, give rise to nontrivial gravitational Casimir effect (GCE). Such GCE \cite{Panella:1993pi} modifies the long-range gravitational interactions by an interaction potential varying with the distance as $r^{-7}$, which is suppressed by the square of the Planck length, in a system that a massive test point particle interacts with a fluctuating mass distribution. A Lifshitz-type formula for the GCE \cite{Quach:2015qwa} at zero temperature in real bodies system gives a gravitational Casimir energy, depending on the frequency of gravitons, which allows us to quantize the gravitational contribution to the Casimir effect. The inspiraling neutron star (NS) binaries, which behave as the tensor sources, are subject to the gravitational interactions and accompany with spin-2 gravitons. So far, the Einstein's general relativity has been widely accepted as a sound theory to describe the dynamics of wide NS binaries, with separation of $R\sim10^9 \rm m$, which move closer and closer in a spiral-in way and may coalesce and merge in the Hubble time. It is the gravitational force to drive the orbital decay, losing orbital binding energy, and emitting gravitational waves (GWs). The spiral-in motion and the decay of orbital separation modify the allowed modes of GWs between two star components and contribute to a shift of gravitational Casimir energy. Because there is no reference for the zero point energy of quantum oscillators, i.e. gravitons, we employ Schwinger's source theory \cite{Schwinger1975, Schwinger1992} and investigate the GCE in wide inspiraling NS binary systems in this work. The shift of gravitational Casimir energy is evaluated, when the orbital separation of two stars decays a small distance $L~(L\ll R)$ ($R$ is the separation of the binary) in radial direction. Firstly, we briefly introduce the preliminary for the scenario of our calculations, in weak-field-limit gravitational source. We also decompose the spin-2 gravitational field into two massless spin-1 scalar fields, by considering the periodicity of the orbital motion of the binary and adopting the Dewitt's approach \cite{DeWitt:1975ys}, in our calculations. The gravitoelectric and gravitomagnetic contributions to the Casimir energy are calculated in the following two sections, respectively. Finally, we estimate the magnitude of GCE and the strength of the corresponding gravitational Casimir force, according to the intrinsic properties and characteristic parameters of the binary system. Because the frequencies of released GWs, in wide inspiraling NS binaries, locate in the low-frequency band of $10^{-4}-1$ Hz, we claim that the gravitational Casimir corrections to the signals would expect to be detected promisingly by space-based observatory LISA/eLISA, with a reduced sensitivity of $10^{-24}$, in the near future. Other possible detections for GCE arising from the inspiraling NS binaries also are discussed in the last part.

Throughout the paper, we use the natural units $c=1$, $G=1$, $\hbar=1$ in the calculations and just write out in the results. The metric signature is defined to be diag(-1,1,1,1). Greek indices $\mu,~\nu$ take values from 0 to 3, while Latin ones $i,~j$ take values from 1 to 3.

\section{Preliminary}
\label{sec:pre}

In wide inspiraling NS binaries, the systems display weak gravitational fields and non-relativistic rotation, which naturally allows us to work in the linearized gravity and assume weak-field approximation to general relativity. In the weak-field limit, the gravitational field of a given NS binary is described by a slight metric perturbation deviating from the flat Minkowskian one, namely,
\begin{equation}
g_{\mu\nu}=\eta_{\mu\nu}+h_{\mu\nu},~~~|h_{\mu\nu}|\ll1.
\end{equation}
The solutions of Einstein's field equation read,
\begin{equation}
ds^2=-(1+\frac{1}{2}\bar{h}^{00})dt^2+2\bar{h}_{0i}dt~dx^{i}+(1-\frac{1}{2}\bar{h}^{00})\delta_{ij}dx^{i}dx^{j},\label{metric}
\end{equation}
where we work in the harmonic gauge, and $\bar{h}_{\mu\nu}\equiv h_{\mu\nu}-\frac{1}{2}\eta_{\mu\nu}h$. The resulting equations resemble those of Maxwell's equations for electromagnetism, referred to as the gravitoelectromagnetism (GEM) \cite{Szekeres:1971ss, Maartens:1997fg, Mashhoon:1999nr, Ruggiero:2002hz}, which is based on the close formal analogy between Newton's law of gravitation and Coulomb's law of electricity. The Newtonian solution of the gravitational field can be alternatively interpreted as a gravitoelectric field, and a rotating mass-current gives rise to a gravitomagnetic field. Accordingly, the Newtonian potential of an NS binary can be analogously written as a gravitoelectric scalar potential, while the orbital rotation of two massive stars causes a gravitomagnetic vector potential. As a consequence, we can write down the gravitoelectric scalar potential and the gravitomagnetic vector potential, respectively,
\begin{equation}
\bar{h}^{00}=4\Phi_{\rm E},~~\bar{h}^{0i}=-2\Phi^i_{\rm M},
\end{equation}
with $\Phi_{\rm E}=-\frac{M}{R}$ ($M$ is the total mass of the binary), and $\Phi^i_{\rm M}\propto\frac{M}{R^2}$. In GEM, analog of mass current to the electric current, the gravitoelectromagnetic fields also satisfy the continuity equation, which is the reduced Lorentz gauge condition in terms of gravitoelectric and gravitomagnetic potentials,
\begin{equation}
\frac{\partial\Phi_{\rm E}}{\partial t}+\frac{1}{2}\nabla_i\Phi_j^M\delta_{ij}=0.\label{{Lorgau}}
\end{equation}
So the total potential of the binary system can be written as
\begin{equation}
\Phi(r)=-\frac{M}{R}+\frac{M}{R^2}r+\mathcal{O}(\frac{M}{R})^3.
\end{equation}
Here, $r$ lies in the range of $0\le r\le L$, and $r/R \sim \mathcal{O}(\frac{M}{R})$, because of the wide separation $R$ of the binary that leads to the orbital decay in the radial direction $r\ll R$, during an observable duration of $m\mathcal{T}$ ($\mathcal{T}$ denotes the orbital period of the NS binary).

According to the Maxwell-like formulation of linearized Einstein field equations, i.e. the gravitoelectromagnetic field equations \cite{Mashhoon:1999nr}, describing the spiral-in dynamics of NS binary, both the gravitoelectric and gravitomagnetic fields should have the form of plane waves,
\begin{equation}
h^{\rm E}_{ij}=\mathcal{E}_{ij}e^{i(\vec{k}\cdot\vec{r}-\omega t)},~~h^{\rm M}_{ij}=\mathcal{M}_{ij}e^{i(\vec{k}\cdot\vec{r}-\omega t)},\label{fields}
\end{equation}
which are transverse waves with two independent polarizations, ``plus'' and ``cross'', each with allowed modes of $(\omega_+,~\omega_{\times})$, respectively. The GWs in linearized Einstein field equations, propagating at the speed of light in all systems, satisfy transverse-traceless (TT) gauge $\partial_ih_{ij}=0$. Only the traceless-tangential waves (or the traceless part of the tangential components of the tensor fields) can be smoothly across the interface \cite{Flanagan:2005yc}, which is known as the smoothness principle \cite{Ingraham:1997dd}. As a consequence, both the gravitoelectric and gravitomagnetic fields are subject to the Casimir-type boundary conditions,
\begin{eqnarray}
&&h_{rr}^{\rm TT}=0,~~h_{\phi z}^{\rm TT}=0,\label{Dbc}\\
&&\partial_rh_{\phi r}^{\rm TT}=\partial_rh_{zr}^{\rm TT}=0.\label{Nbc}
\end{eqnarray}
Here, we use the polar coordinates. The radial component $h_{rr}$ and the cross of angular and perpendicular components ${h_{\phi z}}$ for both fields satisfy Dirichlet boundary conditions, while $h_{\phi r}^{\rm E,M}$ and $h_{zr}^{\rm E,M}$ satisfy Neumann ones.

Following the spirit of DeWitt's approach \cite{DeWitt:1975ys}, the dynamics of physical gravitons in linearized gravity is equivalent to that of a free massless scalar field on a twice interval \cite{Alessio:2020lpk}. Accordingly, we are allowed to make gravitational analogue of the electromagnetic results and decompose the gravitoelectric and gravitomagetic fields into the parts describing their polarizations and a scalar field that is responsible for the dynamical contributions, respectively,
\begin{eqnarray}
&&h^{\rm E}=\sum_kh^{\rm TT+}_{\rm E}\psi_k,~~h^{\rm E}=\sum_kh^{\rm TT \times}_{\rm E}\psi_k,\nonumber\\
&&h^{\rm M}=\sum_kh^{\rm TT+}_{\rm M}\psi_k,~~h^{\rm M}=\sum_kh^{\rm TT \times}_{\rm M}\psi_k.
\end{eqnarray}
The scalar field is independent of the polarizations and gives expression to the allowed modes of the GWs via its dependence of radial coordinate, when the orbital separation decays a distance of $L$ in radial direction after several periods, which consequently contributes to the shift of gravitational Casimir energy. So we turn to evaluate the correction to gravitational Casimir vacuum energy for such a combined massless scalar field, when the binary decay a distance of $L$ in radial separation.

According to Schwinger's source theory \cite{Schwinger1975}, the shift of gravitational vacuum Casimir energy density of an inspiraling NS binary, after $m$ periods $\mathcal{T}$, should be extracted from
\begin{equation}
\langle0_+|0_-\rangle=e^{iW(G)},
\end{equation}
when the separation decays from $R$ to $R-L$. Therefore, we adopt the Schwinger's formalism \cite{Schwinger1992, Toms2007book}, which resorts to the calculations of effective action and study the Casimir energy, without the employment of the concept of zero point energy, and evaluate the effective action $W(G)$ for the gravitoelectromagnetic field resulting from a wide inspiraling NS binary. The effective action, arising from the gravitoelectromagnetism of the sources, contains both the gravitoelectric and gravitomagnetic contributions,
\begin{equation}
W(G)=W_{\rm E}(G)+W_{\rm M}(G).
\end{equation}
The gravitoelectric part $W_{\rm E}(G)$ comes from the Newtonian gravitational interaction, and we refer as the static effect, although it actually isn't stationary. While the gravitomagnetic contribution $W_{\rm M}(G)$ is called the dynamical effect, which is the representation for the fluctuations induced by the gravitomagnetic field.

\section{Static Effect}
\label{sec:static}

In this section, we calculate the contribution from the Newtonian gravitational potential, i.e. the gravitoelectric effect, which yields a diagonal spacetime metric,
\begin{equation}
ds^2=-(1+2\Phi_{\rm E})dt^2+(1-2\Phi_{\rm E})\delta_{ij}dx^idx^j.
\end{equation}
We employ the Schwinger's approach \cite{Schwinger1992} and evaluate the effective action of gravitoelectric part,
\begin{equation}
W_{\rm E}(G)=\lim_{\nu\to0}W_{\rm E}(\nu)=-\frac{i}{2}\int^{\infty}_{s_0}\frac{ds}{s}s^{\nu} \mathrm{Tr} e^{- isH},\label{eaction}
\end{equation}
where the Hamiltonian reads $H=\partial_t^2-\nabla^2=-\omega^2+\vec{p}^2$ and $\omega=\omega_+,~\omega_{\times}$ denotes the summation of all allowed modes. In order to avoid divergence, we take the lower limit of the integration variable $s$ as a very small value $s_0$, and $s_0\to0$ is reserved to the end of the calculations. We will also take the limit $\nu\to0$ at the end of our calculations. The trace in (\ref{eaction}) is evaluated all over the spacetime degrees of freedom,
\begin{equation}
\mathrm{Tr} e^{-isH}=\sum_n\int dtd\vec{r}d\vec{p}_{\perp}d\omega |\langle r|\psi\rangle|^2e^{-is(\vec{p}_{\perp}^2-\omega^2+(\frac{n\pi}{2L})^2)},
\end{equation}
where $\vec{p}_{\perp}$ denote the angular component $p_{\phi}$ and the perpendicular component $p_z$ of the momentum, $n$ represents the quantum numbers of GWs when the orbital decay a distance of $L$ in radial direction. The mode solutions of the combined massless scalar field satisfy the Dirichlet conditions when the binary orbital separation decays from $R$ to $R-L$ \cite{Flanagan:2005yc},
\begin{equation}
\psi(t,\phi, z, r=R) = \psi(t, \phi, z, r=R-L)=0
\end{equation}
The normalized field modes read,
\begin{equation}
\langle r|\psi\rangle^*=\langle\psi|r\rangle=\sqrt{\frac{1}{(1-4\Phi_{\rm E})(2\pi)^3L}}e^{i\omega_{\rm E,n}t}e^{-i\vec{p}_{\perp}\cdot\vec{r}_{\perp}}\sin(\frac{n\pi r}{2L}),
\end{equation}
where $\vec{r}_{\perp}$ denotes the angular component $\phi$ and the perpendicular component $z$ of the coordinates, and the allowed gravitoelectric modes, when the binary orbital separation decays from $R$ to $R-L$, are $\omega^2_{\rm E,n}=(1+4\Phi_{\rm E})(\vec{p}_{\perp}^2+\frac{n^2\pi^2}{4L^2})=(1+4\Phi_{\rm E})(p_{\phi}^2+p_z^2+\frac{n^2\pi^2}{4L^2})$.
Integrating over the spacetime degrees of freedom $dtd\vec{r}d\vec{p}_{\perp}d\omega$ with the duration of $m$ periods $\mathcal{T}$, we obtain the trace in eq. (\ref{eaction}),
\begin{equation}
\mathrm{Tr}e^{-isH}=(1+4\Phi_{\mathrm{E}})\frac{\pi R^2m\mathcal{T}}{\sqrt{i(4\pi s)^3}}\sum_ne^{-is\frac{n^2\pi^2}{4L^2}},\label{etrace}
\end{equation}
when the separation of NS binary system has decayed $L$ in the radial direction in a duration of $m\mathcal{T}$.

Inserting the trace (\ref{etrace}) into eq. (\ref{eaction}), we can write down the gravitoelectric effective action,
\begin{equation}
W_{\mathrm{E}}(\nu)=-\frac{i}{2}\frac{\pi R^2m\mathcal{T}}{\sqrt{i(4\pi)^3}}(1+4\Phi_{\rm E})\int_0^\infty dss^{\nu-\frac{3}{2}-1}\sum_ne^{-is\frac{n^2\pi^2}{4L^2}}.
\end{equation}
Performing the Wick rotation \cite{Cougo-Pinto:1993xqc} and using both the Euler $\Gamma-$function $\Gamma(z)=\int_0^\infty ds~s^{z-1}e^{-s}$ and the Riemann $\zeta-$function $\zeta(z)=\sum_n\frac{1}{n^z}$, we recast the gravitoelectric effective action as,
\begin{equation}
W_{\mathrm{E}}(G)=-\frac{\pi R^2m\mathcal{T}}{2(4\pi)^{3/2}}(-i)^{\nu-2}(\frac{2L}{\pi})^{2\nu-3}\Gamma(2\nu-3)\zeta(2\nu-3)(1+4\Phi_{\rm E}).
\end{equation}
Taking the limit $\nu\to0$ and recalling that $\Gamma(-\frac{3}{2})=\frac{4\sqrt{\pi}}{3}$, $\zeta(-3)=\frac{1}{120}$, we finally have the gravitoelectric contribution to the effective action,
\begin{equation}
W_{\rm E}(G)=(1+4\Phi_{\rm E})\frac{\pi R^2m\mathcal{T}\pi^2}{1440(2L)^3}.\label{GEc}
\end{equation}
With the identification of the energy shift provided by
\begin{equation}
\mathrm{E}^{\rm E}_{\rm Cas}=W_{\rm E}(G)t
\end{equation}
in the duration time $t$, we get the gravitoelectric-induced shift of gravitational Casimir energy, when an NS binary orbits a duration of $t=m\mathcal{T}$ and the orbital separation decays a distance of $L$,
\begin{equation}
\mathrm{E}^{\rm E}_{\rm Cas}=\frac{W_{\rm E}(G)}{m\mathcal{T}}=-\frac{\pi R^2\hbar c\pi^2}{1440(2L)^3}(1+4\Phi_{\rm E}).
\end{equation}
The corresponding gravitoelectric Casimir energy density is then given by
\begin{equation}
\mathcal{E}^{\rm E}_{\rm Cas}=-\frac{\hbar c\pi^2}{720(2L)^4}(1+4\Phi_{\rm E}).
\end{equation}

\section{Dynamical Effect}\label{sec:dyn}

We now move to adopt the above technique to the evaluation of the gravitomagnetic correction to the gravitational Casimir energy density, during the spiral-in orbital motion. By taking the gravitomagnetic contributions into account, the spacetime metric then reads
\begin{equation}
ds^2=-(1+2\Phi_{\rm E}+2\Phi_{\rm M}r)dt^2+(1-2\Phi_{\rm E}-2\Phi_{\rm M}r)\delta_{ij}dx^idx^j,
\end{equation}
and the equation of motion can be written down,
\begin{equation}
-(1-4\Phi_{\rm E}-4\Phi_{\rm M}r)\partial_t^2\psi+\delta_{ij}\partial x_i\partial x_j\psi=0.
\end{equation}
We are interested in the mode solutions describing the combined single massless scalar field, when the binary orbital separation decays from $R$ to $R-L$,
\begin{equation}
\psi=\frac{1}{\sqrt{(1-4\Phi_{\rm E}-4\Phi_{\rm M}(2L))(2\pi)^3\omega_{\rm M,n}(2L)}}e^{-i\omega_{\mathrm{M,n}}t}e^{i\vec{p}_{\perp}\cdot\vec{r}_{\perp}}\sin(\frac{n\pi}{2L}r),
\end{equation}
where the GW frequencies of allowed GEM modes are $\omega^2_{\rm M,n}=(1+4\Phi_{\rm E}+4\Phi_{\rm M}(2L))(\vec{p}_{\perp}^2+\frac{n^2\pi^2}{4L^2})=(1+4\Phi_{\rm E}+4\Phi_{\rm M}(2L))(p_{\phi}^2+p^2_z+\frac{n^2\pi^2}{4L^2})$.

Then we are allowed to calculate the gravitomagnetic effective action,
\begin{eqnarray}
W_{\rm M}(G)&=&\lim_{\nu\to0}W_{\rm M}(\nu)\nonumber\\
&=&-\frac{i}{2}\int_0^{\infty}dss^{\nu-1}\mathrm{Tr}[4i\Phi_{\rm M}s(r+is\partial_r)\nabla^2e^{-isH}].\label{GMaction}
\end{eqnarray}
Thereinto, the trace can be get by integrating over the spacetime during the duration of $m\mathcal{T}$ inspiral,
\begin{eqnarray}
&&\mathrm{Tr}[4i\Phi_{\rm M}s(r+is\partial_r)\nabla^2e^{-isH}]\nonumber\\
&=&4i\Phi_{\rm M}s\int d\vec{r}dt\sum_n\int d\vec{p}_Ad\omega|\langle r|\psi\rangle|^2(r+is\partial_r)(-\vec{p}^2_A-\frac{n^2\pi^2}{4L^2})\sin^2\frac{n\pi r}{2L}e^{-is(-\omega^2+\vec{p^2_A+\frac{n^2\pi^2}{4L^2}})}\nonumber\\
&=&\frac{4i\Phi_{\rm M}s}{(2\pi)^{3/2}(2L)}\sum_n\int d\vec{r}dtd\vec{p}_Ad\omega\sin^2(\frac{n\pi}{2L})(r+is\partial_r)(-\vec{p}^2_A-\frac{n^2\pi^2}{4L^2})\sin^2\frac{n\pi r}{2L}e^{-is(-\omega^2+\vec{p}^2_A+\frac{n^2\pi^2}{4L^2})}\nonumber\\
&=&\frac{i8\Phi_{\rm M}\pi R^2m\mathcal{T}\sqrt{\pi}}{(2\pi)^2L\sqrt{-i}}s^{\frac{1}{2}}\int_0^{2L}dr\sum_n\sin^2\frac{n\pi r}{2L}e^{-is\frac{n^2\pi^2}{4L^2}}(-\frac{\pi r}{s^2}+\frac{\pi r}{is}\frac{n^2\pi^2}{4L^2}+\frac{\pi}{s}\frac{n\pi}{2L}-\frac{\pi}{i}\frac{n^3\pi^3}{8L^3}).
\end{eqnarray}
Substituting the above result into eq. (\ref{GMaction}), we evaluate the gravitomagnetic contribution to the effective action,
\begin{eqnarray}
W_{\rm M}(\nu)&=&\frac{4\Phi_{\rm M}\pi R^2m\mathcal{T}\sqrt{\pi}}{(2\pi)^3L\sqrt{-i}}\sum_n\int_0^{\infty}dse^{-is\frac{n^2\pi^2}{4L^2}}[-\pi L^2(s^{\nu-\frac{3}{2}-1}+i\frac{n^2\pi^2}{4L^2}s^{\nu-\frac{1}{2}-1})\nonumber\\
&&+\pi L\frac{n\pi}{2L}(s^{\nu-\frac{1}{2}-1}+i\frac{n^2\pi^2}{4L^2}s^{\nu+\frac{1}{2}-1})].\label{wm}
\end{eqnarray}
The contributions all come from the terms in the first line of eq. (\ref{wm}), while the terms in the second line of eq. (\ref{wm}), proportional to $\zeta(-2)=0$, give no any contribution. By taking the limit $\nu=0$, the gravitomagnetic corrected effective action is written as
\begin{eqnarray}
W_{\rm M}(G)=\frac{4\Phi_{\rm M}\pi R^2m\mathcal{T}\sqrt{\pi}}{(2\pi)^3L\sqrt{-i}}\frac{\pi^4}{L}[-(-i)^{-\frac{3}{2}}\Gamma(-\frac{3}{2})\zeta(-3)+(-i)^{\frac{1}{2}}\Gamma(-\frac{1}{2})\zeta(-3)].
\end{eqnarray}
Considering that $\Gamma(-\frac{3}{2})=\frac{4\sqrt{\pi}}{3}$, $\Gamma(-\frac{1}{2})=-2\sqrt{\pi}$,  $\zeta(-3)=\frac{1}{120}$, we finally obtain the effective action of the gravitomagnetic part,
\begin{equation}
W_{\rm M}(G)=\frac{\pi R^2m\mathcal{T}\pi^2}{1440(2L)^3}\Phi_{\rm M}(2L),\label{GMc}
\end{equation}

Combined eq. (\ref{GEc}) and eq. (\ref{GMc}), we can write down the full effective action of GEM in inspiraling NS binaries,
\begin{eqnarray}
W(G) = W_{\rm E}(G) +W_{\rm M}(G) &=& (1+4\Phi_{\rm E})\frac{\pi R^2m\mathcal{T}\pi^2}{1440(2L)^3} + \frac{\pi R^2m\mathcal{T}\pi^2}{1440(2L)^3}\Phi_{\rm M}(2L)\nonumber\\
&=& \frac{\pi R^2m\mathcal{T}\pi^2}{1440(2L)^3}(1+4\Phi_{\rm E}+2\Phi_{\rm M}L).
\end{eqnarray}
The corresponding shift of gravitational Casimir energy density, when the binary orbital separation decays a distance of $L$ in a duration of $m$ periods $\mathcal{T}$, is then
\begin{equation}
\mathcal{E}_{\rm Cas}=-\frac{\hbar c\pi^2}{720(2L)^4}(1+4\Phi_{\rm E}+2\Phi_{\rm M}L).\label{tGCE}
\end{equation}

\section{Summary and discussions}\label{sec:dis}

In an inspiraling NS binary, two stars in the system gravitationally interact and orbit with each other in a spiral-in way, losing orbital energy and releasing GWs. Based upon the Schwinger's source theory, we study the shift of gravitational Casimir energy for the massless spin-2 gravitons, released from the inspiraling processes of wide NS binaries, with separation of $R\sim10^9$ m, when the orbital separation decays a distance of $L$ in radial direction after $m$ orbital periods $\mathcal{T}$. The spiral-in orbital motion of two massive stars in the NS binary gives rise to a gravitoelectromagnetic field, which allows us to make analogy to the Casimir effect of electromagnetic field. We compute both gravitoelectric and gravitomagnetic corrections to the gravitational Casimir energy during the spiral-in orbital motion, in the weak-field-limit approximation. It is found that a net GCE due to the appearance of GEM if the orbital separation of an NS binary decays a nontrivial distance in radial direction after $m\mathcal{T}$.

Subject to the Casimir-type boundary conditions (\ref{Dbc}) and (\ref{Nbc}), the GCE behaves as a manifestation of the quantization of released GWs, or gravitons, when the orbital separation decays a nontrivial distance of $L$ in a duration of $m$ periods $\mathcal{T}$. Associated with the shift of gravitational Casimir energy, a corresponding gravitational Casimir force subsequently appears and may have influence on the orbital motion. In light of the gravitational Casimir energy when the binary orbital separation decays from $R$ to $R-L$, we immediately get a consequence that it gives rise to an attractive force. The gravitational Casimir force density can be evaluated as
\begin{equation}
F_{\rm Cas}=-\frac{\partial \mathcal{E}_{\rm cas}}{\partial L}=-\frac{\pi^2\hbar c}{240(2L)^4}(1+4\Phi_{\rm E}+\frac{4}{3}\Phi_{\rm M}L).
\end{equation}
That is to say, the inspiraling NS binary is subject to an attractively gravitational Casimir force, in addition to the gravitational interaction. Even though the quantum nature of such an attractive force, it arises from the GEM corrections to the general gravitational fields, due to the relativistic massive NS current. If the orbital separation decays a distance of $1\rm m$, the binary system may suffer from a gravitational Casimir force of an order of $10^{-21} \rm N$. The gravitoelectric correction leads to a contribution of an order of $\sim\frac{M}{R}$ (the total mass of the system is about $M\sim$ several solar mass), which results in a nontrivial correction to the additional attractive force. While the gravitomagnetic part, with the order of $\sim\frac{M}{R^2}$, brings about a relatively insignificant contribution, which is consistent with the result by the previous work about the influence of a gravitomagnetic field on the vacuum energy of a scalar field in a rigid Casimir cavity \cite{Sorge:2009zz}.

\begin{figure*}
\includegraphics[bb=300 50 660 600,width=8cm]{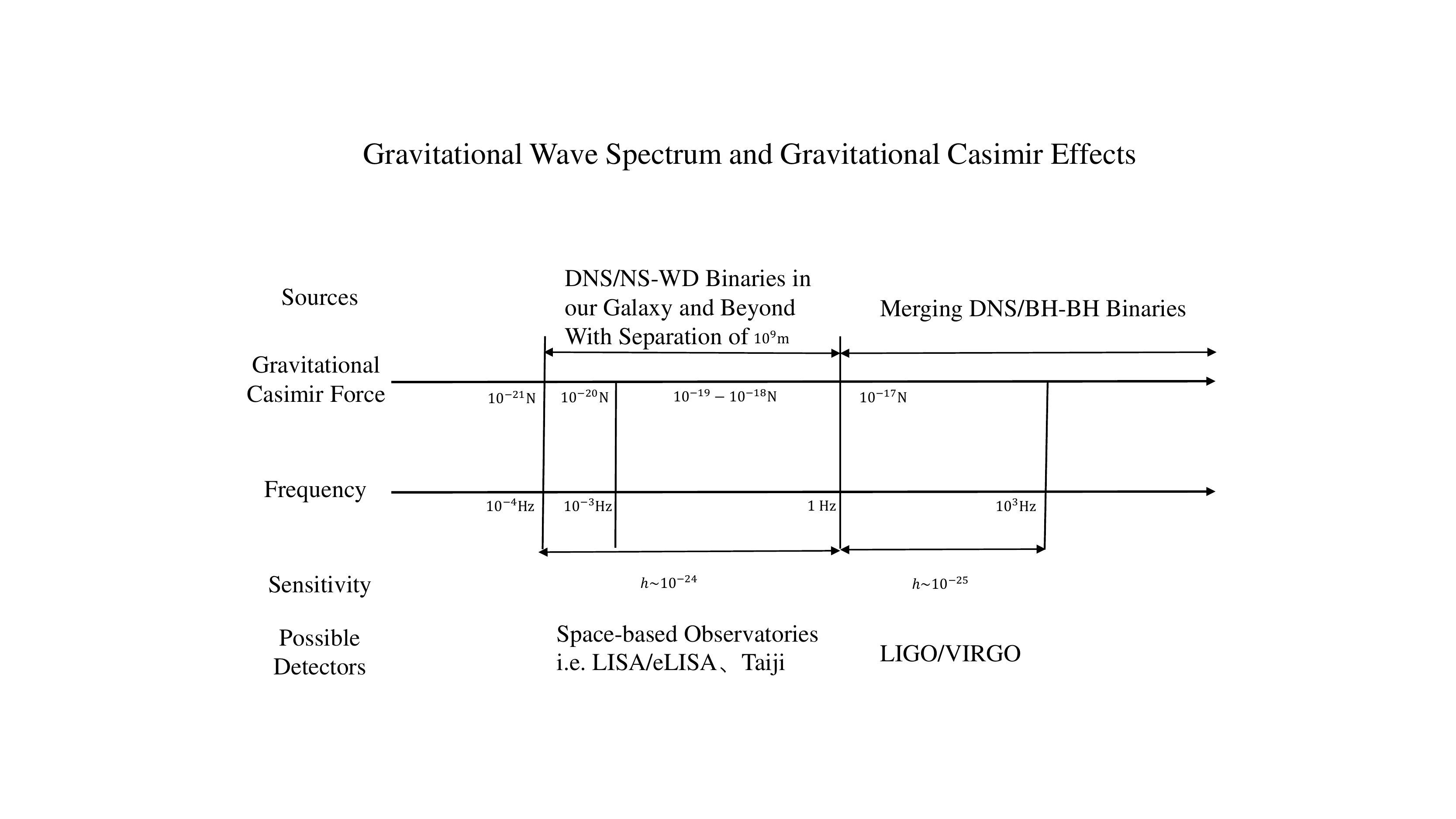}
\caption{The gravitational Casimir force in inspiraling neutron star binaries corresponding to respective gravitational waves spectrum, and the possible detectors with necessary sensitivities.}\label{fig:GCEspectrum}
\end{figure*}

By considering the order of $10^{-21} \rm N$, it turns out to be only compatible with the extremely sensitive force detectors in order to detect such a GEM-induced attractive gravitational Casimir force, although in a considerable observing time. Even the frustrating result in the sense of measuring the GEM corrections to the gravitational Casimir energy because of its extremely smallness but the non-trivialness, we expect the GCE in inspiraling nd low-frequency GWs sources can be detected if the modulation signal is higher than the sensitivity of the detector. In wide inspiraling NS binaries, the characteristic frequency of GWs is $10^{-4}-1 \rm Hz$. The quantum corrections to the signals are estimated as of the order of $h\sim10^{-24}$, corresponding to a force of magnitude $\sim10^{-20} \rm N$ for a typical frequency of $10^{-3} \rm Hz$. According to the intrinsic properties and characteristic parameters of NS binary systems, we estimate the order of magnitude of gravitational Casimir force, corresponding to different frequency bands of GWs, and list the the possible detectors with necessary sensitivities, which are plotted in Figure \ref{fig:GCEspectrum}. The gravitational Casimir force and the gravitoelectric correction in wide inspiraling NS binaries are very promising to be detected with improved sensitivity of space-based GW observatories in the near future. If the sensitivity of LISA/eLISA can be reduced to $10^{-24}$, it will be the promising tool to detect the GCE on wide inspiraling NS binaries. While for the GW signals of order of $h\sim10^{-25}$ in coalescing phase, the corresponding forces of magnitude is about $\sim10^{-17} \rm N$ at LIGO/VIRGO frequency band of few tens of Hz. However, it may need a higher sensitivity than $10^{-25}$ for the ground-based interferometries to detect the quantum effects on high-frequency GWs, by considering both more interferences to the detection of signals than the space-based observatories and the stronger relativistic effects during the merger phase because of the regime of stronger gravity, which may hide the quantum ones. As a summary, we expect that the LISA/eLISA and Taiji GW detectors with sensitivity improvements \cite{Calloni:2001mb} would give the powerful tool to detect such effects. The GCE from high-frequency GWs sources, such as the merging black-hole/NS binaries during coalescing phase, also can be expected to be detected by LIGO/VIRGO with improved sensitivity higher than $10^{-25}$. Apart from the possible direct GW detections, we also suggest the cosmological observations, combining with pulsar timing array, for the GCE, by considering the effects of GEM on large scales \cite{Caldwell:2002im, Brevik:2000zb, Mostepanenko:2000kn}.

\begin{acknowledgments}
This work is supported by the Guangxi Natural Science Foundation Program (Grant no. 007151339018) and by Guangxi Science and Technology Foundation and Talent Special (Grant no. 111252047014).
\end{acknowledgments}

\end{document}